\documentclass[aps,prl,showpacs]{revtex4}
\usepackage{graphicx}
\usepackage{amsfonts,amssymb,amsmath}
\usepackage{bm}

\begin{document}

\title{A Unified Treatment of Quasi-Exactly Solvable Potentials II: Eckart Type Potentials}
\date{\today}
 
\author{Ramazan Ko\c{c}}
\email{koc@gantep.edu.tr}
\affiliation{Department of Physics, Faculty of Engineering 
University of Gaziantep,  27310 Gaziantep, Turkey}
\author{Mehmet Koca}
\email{kocam@squ.edu.om}
\affiliation{Department of Physics, College of Science,
Sultan Qaboos University, PO Box 36  \\
Al-Khod 123, Sultanete of Oman}

\begin{abstract}
This work continues to study the set of quasi exactly solvable potentials
related to the Eckart, Hult\'{e}n, Rosen-Morse, Coulomb and the harmonic
oscillator potentials. We solve the Schr\"{o}dinger equation for each
potential and obtain the eigenvalues and eigenstates in terms of the
orthogonal polynomials. We present a unified approach to obtain the
potentials from each other by suitable change of variables.
\end{abstract}
\maketitle
\section{Introduction}

The quasi exactly solvable(QES) quantum mechanical problems are such that
several eigenstates can be found explicitly. They occupy an intermediate
place between exactly solvable and non-solvable potentials\cite{turb1}. A
general treatment of the quasi-exact solvability has been introduced by
Turbiner and Ushveridze\cite{turb2} and they were classified by
Gonzalez-Lopez, Kamran and Olver\cite{gonz1, gonz2}. \ A number of papers
deal with the several aspects of the QES potentials\cite{deber1, kuliy,
brihaye, znojil}. One approach to determine the eigenfunctions of the QES
Schr\"{o}dinger equation is the orthogonal polynomials which appear in the
coefficients of the eigenfunctions. In a previous paper we have obtained
eigenvalues and eigenfunctions of the Schr\"{o}dinger equation with QES
potentials associated with the P\"{o}schle-Teller, the generalized P\"{o}%
schle-Teller, the PT symmetric Scarf and the harmonic oscillator potentials.

Much less attention has been paid to the QES potentials related to the
Eckart, Hult\'{e}n and Rosen-Morse potentials. We construct a general
algorithm generating the eigenstates as well as eigenvalues of the
corresponding potentials by employing the method suggested by Bender and
Dunne\cite{Bender1,Bender2}. They have showed that there is a
correspondence between QES models in quantum mechanics and the set of
orthogonal polynomials. Systematic studies of these polynomials have been
given in the papers\cite{gonz3} and\cite{krajew}.

In this paper we extend the method to obtain the eigenstates and eigenvalues
of the Schr\"{o}dinger equation with QES Eckart potential and then we
transform it to the QES Hult\'{e}n and the QES Rosen-Morse potentials by a
linear transformation on the coordinate. We also give a procedure to convert
the QES Eckart and the QES Rosen-Morse potentials to the QES Coulomb and the
QES harmonic oscillator potentials, respectively.

In section 2, the QES Eckart potential is constructed by using the $sl(2,R)$
algebra. The eigenvalues and eigenfunctions of the Schr\"{o}dinger equation
are obtained in terms of the orthogonal polynomials. Sections 3 and 4 deal
with the transformation of QES Eckart potential to the Hult\'{e}n and
Rosen-Morse potentials. The relation between the QES Eckart potential and
the QES Coulomb potential is discussed in section 5. We also show that a
specific form of the QES Coulomb potential corresponds to the Coulomb
correlation problem for a system of two electrons in an external oscillator
potential. The results obtained here are compared with the results given in
the paper\cite{turb1}. In section 6, the QES Rosen-Morse potential is
converted to the anharmonic oscillator potential. Finally, we discuss the
validity of our results in section 7.

\section{QES Eckart Potential}

The QES potentials have been constructed by employing a variety of
approaches \cite{turb2, gonz1, turb4}. In this section we obtain the QES
Eckart potential, which can be generated by using the Lie algebraic method,
widely used in the literature. Let us consider the second order differential
equation which can be obtained by using linear and bilinear combinations of
the operators of the $sl(2,R)$.The QES Eckart potential given in this paper can be 
obtained by transforming
the following linear and bilinear combination of the operators of the $%
sl(2,R)$ Lie algebra,%
\begin{eqnarray*}
&&2J_{-}J_{+}+2J_{-}J_{0}-(\lambda +2(2L+j+3))J_{-}+\lambda J_{0}-qJ_{+} \\
&&+(\lambda (L+j+1)+(2(L(L+3)+2j(L+1)-A)
\end{eqnarray*}%
in the form of Schrodinger equation. The generators of the Lie algebra
satisfy the standard differential realizatin of the$\quad sl(2,R)$ \cite%
{turb2, gonz1, turb4}, with standard realization,%
\begin{align}
& {2z(1-z)\frac{{d^{2}\Re _{j}(z)}}{{dz^{2}}}+\left( {4(jz-L-j-1)+\lambda
(z-1)+2qz^{2}}\right) \frac{{d\Re _{j}(z)}}{{dz}}}  \notag \\
& +{\left( {\lambda (L+1)+2L(L+3)-A+4j(L-qz+1)+4}\right) \Re _{j}(z)}=0
\label{eq:1}
\end{align}%
which possesses the polynomial solution. Here $L,q,A$ and $\lambda $ are
constants, and $j=0,1/2,1,\cdots $. The function $\Re _{j}(z)$ forms a basis
function for the $sl(2,R)$ algebra and preserve the space of polynomials of
order $2j$. The differential equation (\ref{eq:1}) is exactly solvable under
the condition $q=0$. In order to generate the QES quantum mechanical
potentials, two-step procedure can be employed to transform (1) into the Schr%
\"{o}dinger equation. We introduce a new variable,%
\begin{equation}
z=e^{-2\alpha x}  \label{eq:2}
\end{equation}%
as well as define a new function, 
\begin{eqnarray}
\psi (x) &=&\left( {e^{-2\alpha x}-1}\right) ^{L-{\frac{q}{2}}}\left( {%
1-e^{2\alpha x}}\right) \times  \label{eq:3} \\
&&\exp \left( {\frac{1}{2}\alpha x(4L+4j+2+\lambda )-\frac{1}{2}qe^{-2\alpha
x}}\right) \Re _{j}\left( {e^{-2\alpha x}}\right)  \notag
\end{eqnarray}%
to eliminate the first order differential term in (\ref{eq:1}). The equation
is converted to the Schr\"{o}dinger equation:%
\begin{equation}
-\frac{{d^{2}\psi (x)}}{{dx^{2}}}+(V(x)-E)\psi (x)=0.  \label{eq:4}
\end{equation}%
Here the potential $V(x)$ is in the form:%
\begin{eqnarray}
V(x) &=&\left[ L(L+1)\alpha ^{2}\csc h^{2}\alpha x+A\alpha ^{2}\coth \alpha x%
\right]  \notag \\
&&+\frac{q\alpha ^{2}}{\left( e^{2\alpha x}-1\right) ^{2}}\left(
qe^{-4\alpha x}+(\lambda -4j)e^{-2\alpha x}-(4L-4j+\lambda +2)\right) .
\label{eq:5}
\end{eqnarray}%
The energy eigenvalues for this potential can be determined as%
\begin{equation}
E=\left( A-(2L+\lambda /2+4j+3)^{2}\right) \alpha ^{2}.  \label{eq:6}
\end{equation}%
The term given in [$\cdots $] in(\ref{eq:5}) is the Eckart potential. When
we choose $q=0$ and,%
\begin{equation}
\lambda =\frac{A}{L+m+1}-2(L+2j-m+2),  \label{eq:7}
\end{equation}%
then we obtain the exact solution of the Schr\"{o}dinger equation where the
eigenvalues (\ref{eq:6}) of the Schr\"{o}dinger equation can be expressed in
the closed form%
\begin{equation}
E_{e}=\frac{\alpha ^{2}}{4}\left( \left( \frac{A}{L+m+1}\right)
^{2}+4(L+m+1)^{2}\right)  \label{eq:8}
\end{equation}%
and the eigenfunctions can be expressed in terms of the Jacobi polynomials.
As it was stated before, for the differential equation (\ref{eq:1}) to
possess polynomial solutions for non negative integer and half-integer
values of $j$, ${\Re _{j}(z)}$ can be written as a polynomial of degree $2j:$%
\begin{equation}
\Re _{j}(z)=\sum\limits_{m=0}^{2j}{a_{m}z^{m}.}  \label{eq:9}
\end{equation}%
The coefficient $a_{m}$ is related to the parameters of the potential. Let
us redefine the coefficient $a_{m}$ such that the parameter $\lambda $
factors out from the other parameters:%
\begin{equation}
a_{m}=f_{m}(L,A,q)P_{m}(\lambda ).  \label{eq:10}
\end{equation}%
The coefficient, $f_{m}(A,L,q),$ is independent of $\lambda $ and plays the
role of the normalization constant. After a straightforward calculation one
can obtain a three-term recurrence relation for the polynomial $%
P_{m}(\lambda ):$%
\begin{align}
& 2q(2j-m)P_{m+1}(\lambda )-\left( {(\lambda +2(L+2j-m+2))(L+m+1)-A}\right)
P_{m}(\lambda )  \notag \\
& +m(4L+4j-2m+6+\lambda )P_{m-1}(\lambda )=0  \label{eq:11}
\end{align}%
with the initial condition $P_{0}(\lambda )=1$. The polynomial $P_{2j+1}$
vanishes and the roots $\lambda $ belong to the eigenvalues of the Schr\"{o}%
dinger equation. Analytical solution of the recurrence relation(\ref{eq:11}%
), for $\lambda ,$ depends on the values of the parameters $A,L$ and $q$.
But, in general, for $j\geq 2$ the values of $\lambda $ can be determined
numerically.

\section{QES Hult\'{e}n potential}

In this section we discuss the transformation of QES Eckart potential to the
QES Hult\'{e}n potential, the exactly solvable form of which has been widely
discussed in the literature\cite{znojil4, arin}. The QES Eckart potential
can be converted to the QES Hult\'{e}n potential by replacing $x\rightarrow
x/2.$ Under these conditions, we obtain QES Hult\'{e}n potential:%
\begin{eqnarray}
V(x) &=&\frac{\alpha ^{2}}{2}\left( 2L(L+1)+A\right) \frac{e^{-\alpha x}}{%
1-e^{-\alpha x}}+L(L+1)A\left( \frac{\alpha e^{-\alpha x}}{1-e^{-\alpha x}}%
\right) ^{2}  \notag \\
&&+\frac{q\alpha ^{2}}{4\left( e^{2\alpha x}-1\right) ^{2}}\left(
qe^{-2\alpha x}+(\lambda -4j)e^{-\alpha x}-(4L-4j+\lambda +2)\right)
\label{eq:12}
\end{eqnarray}%
The energy values of the Schr\"{o}dinger equation with QES Hult\'{e}n
potential(\ref{eq:12}) is given by 
\begin{equation}
E^{\prime }=\frac{1}{4}(E_{e}+A\alpha ^{2})  \label{eq:13}
\end{equation}%
where $E$ is the energy eigenvalues of the Schr\"{o}dinger equation with QES
Eckart potential of(\ref{eq:6}) . The corresponding eigenfunction is then
given by%
\begin{equation}
\psi _{_{HL}}(x)=\psi (x/2)  \label{eq:14}
\end{equation}%
where $\psi (x)$ is the wave function of the Schr\"{o}dinger equation with
the QES Eckart potential. It is obvious that the corresponding polynomial $%
P_{m}(\lambda )$ satisfies the same recurrence relation of \ (\ref{eq:11}).
Under the conditions that $q=0$ and $\lambda $ takes the value of(\ref{eq:7}%
), the Hult\'{e}n potential is exactly solvable and the eigenvalues of the
Schr\"{o}dinger equation read%
\begin{equation}
E_{e}^{\prime }=\frac{1}{4}(E_{e}+A\alpha ^{2})
\end{equation}%
where $E_{e}$ is given in(\ref{eq:8}).

\section{QES Rosen-Morse potential}

The QES Eckart potential can be transformed to the Rosen- Morse potential by
shifting the coordinate $x\rightarrow x+i\pi /4\alpha .$ Then the
corresponding potential is given by%
\begin{eqnarray}
V(x) &=&\left[ -L(L+1)\alpha ^{2}\sec h^{2}\alpha x+A\alpha ^{2}\tan \alpha x%
\right]  \notag \\
&&+\frac{q\alpha ^{2}}{\left( e^{2\alpha x}-1\right) ^{2}}\left(
qe^{-4\alpha x}+(\lambda -4j)e^{-2\alpha x}-(4L-4j+\lambda +2)\right)
\label{eq:15}
\end{eqnarray}%
The energy eigenvalues are the same as in the QES Eckart potential. The
corresponding wave function is $\psi _{_{RM}}(x)=\psi (-x).$The polynomials
in the coefficient of the eigenfunction of the QES Rosen- Morse potential
obey the same recurrence relation given in(\ref{eq:11}).

So far we have discussed the QES potentials associated with the Eckart, Hult%
\'{e}n and Rosen- Morse potentials and obtained the relations among them.
Some of these relations were previously discussed in the literature but not
all. In the following sections we give a procedure to convert the QES Eckart
and QES Rosen- Morse potentials to the QES Coulomb and QES harmonic
oscillator potentials, respectively.

\section{QES Coulomb potential}

The QES Eckart potential(\ref{eq:5}) can be transformed to the QES Coulomb
potential by a suitable change of parameters. Let us redefine the parameters
by%
\begin{equation}
q\rightarrow \frac{q}{{2\alpha ^{2}}},\quad L\rightarrow \ell +\frac{q}{{%
4\alpha ^{2}}},\quad \lambda \rightarrow 2+4j+\frac{{a+\varepsilon }}{\alpha 
}-\frac{{2q}}{{\alpha ^{2}}},\quad A\rightarrow \frac{a}{\alpha }-\frac{{%
q\ell }}{{\alpha ^{2}}}+\frac{{4aq}}{{\alpha ^{3}}}-\frac{{3q^{2}}}{{8\alpha
^{4}}}  \label{eq:16}
\end{equation}%
and substitute them in (\ref{eq:5}) and (\ref{eq:6}); when we take the limit
of the $\left( V(x)-E\right) $, as $\alpha \rightarrow 0$, the potential is
transformed to the perturbed Coulomb potential:%
\begin{equation}
V(x)=\frac{{\ell (\ell +1)}}{{x^{2}}}+\frac{a}{x}-q\left( {\varepsilon +a}%
\right) x+q^{2}x^{2}.  \label{eq:17}
\end{equation}%
This relation indicates that the QES Eckart potential and QES Coulomb
potential possesses similar behavior in a certain range of variable $x$. The
ground state wave function of the perturbed Coulomb potential is given by%
\begin{equation}
\psi (x)=x^{\ell +1}\exp \left( {\frac{1}{2}x\left( {\varepsilon +a-qx}%
\right) }\right)  \label{eq:18}
\end{equation}%
and the energy eigenvalue reads as%
\begin{equation}
E=-\frac{{(\varepsilon +a)^{2}}}{4}+2q\left( {\ell +2j+\frac{3}{2}}\right) .
\label{eq:19}
\end{equation}%
The parameter ${\varepsilon }$ is obtained from the recurrence relation%
\begin{equation}
2q(2j-m)P_{m+1}(\varepsilon )+\left( {(\varepsilon +a)(\ell +m+1)+a}\right)
P_{m}({\varepsilon })+m\left( {2\ell +m+1}\right) P_{m-1}({\varepsilon })=0.
\label{eq:20}
\end{equation}%
When $q=0$, the potential is exactly solvable and from(\ref{eq:20}) we obtain%
\begin{equation}
\varepsilon =-a\left( {1+\frac{1}{{\ell +m+1}}}\right) .  \label{eq:21}
\end{equation}%
Then energy eigenvalue(\ref{eq:19}) takes the form%
\begin{equation}
E=\left( {\frac{{a/2}}{{\ell +m+1}}}\right) ^{2}.  \label{eq:22}
\end{equation}%
With the substitution(\ref{eq:17}) and $\alpha \rightarrow 0$ one can
transform the QES Hult\'{e}n potential into the perturbed Coulomb potential.
We note here that, with a special choice of ${\varepsilon }=-a$ the QES
Coulomb potential corresponds to the potential describing a two-electron
system in an external oscillator potential\cite{turb3, taut} where the
eigenvalue of the Schr\"{o}dinger equation reads%
\begin{equation}
E=2q\left( {\ell +2j+\frac{3}{2}}\right)  \label{eq:30}
\end{equation}%
With the substitution $\varepsilon =-a$ in(\ref{eq:20})the recurrence
relation simplifies to%
\begin{equation}
2q(2j-m)P_{m+1}(\varepsilon )-\varepsilon P_{m}({\varepsilon })+m\left( {%
2\ell +m+1}\right) P_{m-1}({\varepsilon })=0  \label{eq:31}
\end{equation}%
The parameter $a(\varepsilon =-a)$ is the effective charge of the
inter-electronic interaction and it can be changed by the atomic number Z.
The effective charge can be obtained by solving the recurrence relation. The
first four of this relation is given by%
\begin{eqnarray}
P_{1}({\varepsilon }) &=&{\varepsilon }  \notag \\
P_{2}({\varepsilon }) &=&{\varepsilon }^{2}-8qj(\ell +1)  \notag \\
P_{3}({\varepsilon }) &=&{\varepsilon }\left[ {\varepsilon }^{2}+4q(2\ell
+3-2j(3\ell +4))\right]  \notag \\
P_{4}({\varepsilon }) &=&{\varepsilon }^{4}+4q(8\ell +15-4j(4\ell +5)){%
\varepsilon }^{2}+192j(j-1)(\ell ^{2}+3\ell +2)q^{2}
\end{eqnarray}%
The roots of the polynomial $P_{2j+1}({\varepsilon )=0}$ can be written for
a few values of j,%
\begin{eqnarray}
j=0,\quad {\varepsilon } &{=}&{0}  \notag \\
j=\frac{1}{2},\quad {\varepsilon } &{=}&{\pm 2}\sqrt{q(\ell +1)}  \notag \\
j=1,\quad {\varepsilon } &{=}&\left\{ {\pm 2}\sqrt{q(4\ell +5)},\quad
0\right\}  \notag \\
j=\frac{3}{2},\quad {\varepsilon } &{=}&\left\{ \pm \sqrt{2q\left( 5(2\ell
+3)\pm \sqrt{153+64\ell (\ell +3)}\right) }\right\} .
\end{eqnarray}%
As noted in the paper\cite{turb1}, there exists equal number of positive and
negative eigenvalues. This implies that electron-electron correlation energy
is related with the electron-positron correlation energy.

\section{QES Harmonic oscillator potential}

In order to obtain the anharmonic oscillator potential from the Rosen- Morse
potential we redefine the parameters%
\begin{eqnarray}
q &\rightarrow &\frac{q}{{2\alpha ^{3}}},\quad L\rightarrow -\frac{\ell }{{%
4\alpha }}+\frac{a}{{2\alpha ^{2}}}-\frac{{3q}}{{2\alpha ^{3}}}  \notag \\
\lambda &\rightarrow &\varepsilon +\frac{{3\ell }}{{2\alpha }}-\frac{a}{{%
\alpha ^{2}}}+\frac{{2q}}{{\alpha ^{3}}}  \notag \\
A &\rightarrow &\frac{{2a\ell -q(3\varepsilon +4j+10)}}{{4\alpha ^{3}}}-%
\frac{{7q\ell }}{{8\alpha ^{4}}}+\frac{{qa}}{{4\alpha ^{5}}}-\frac{{3q^{2}}}{%
{8\alpha ^{6}}}.  \label{eq:23}
\end{eqnarray}%
Substituting (\ref{eq:23}) into (\ref{eq:6}) and (\ref{eq:14}), taking the
limit of $\left( V(x)-E\right) $ when $\alpha \rightarrow 0$, we obtain the
potential,%
\begin{equation}
V(x)=\left( {\frac{{a\ell }}{2}-2q(1+2j)}\right) x+\left( {\frac{{a^{2}}}{4}%
-q\ell }\right) x^{2}-qax^{3}+q^{2}x^{4}.  \label{24}
\end{equation}%
The ground state wave function is given by%
\begin{equation}
\psi (x)=\exp \left( {\frac{\ell }{2}x+\frac{a}{4}x^{2}-\frac{q}{3}x^{3}}%
\right)  \label{25}
\end{equation}%
and the energy is obtained as%
\begin{equation}
E=\frac{1}{2}\left( {\varepsilon +4j+5}\right) a.  \label{26}
\end{equation}%
As noted before ${\varepsilon }$ is determined from the recurrence relation%
\begin{gather}
2q(m-2j)P_{m+1}(\varepsilon )+\frac{1}{4}\left( {2a(\varepsilon
+4j-2m+4)-\ell ^{2}}\right) P_{m}(\varepsilon )  \notag \\
-m\ell P_{m-1}(\varepsilon )-m(m-1)P_{m-2}(\varepsilon )=0
\end{gather}%
This transformation shows that the QES Rosen-Morse and the anharmonic
oscillator potentials have similar behavior in a certain range of variable $%
x $.

\section{Conclusion}

In order to determine eigenstates and eigenvalues of the various potentials
we have developed a method based on the orthogonal polynomials. We have
obtained the recurrence relations of the polynomials, which represent the
eigenstates of the corresponding potentials. It is interesting to observe
that the polynomial of energy occuring in the eigenfunctions of the QES
Eckart, QES Hult\'{e}n and the QES Rosen-Morse potentials satisfy the same
recurrence relation(\ref{eq:11}). We have also have proven that the QES
Eckart and QES Rosen-Morse potentials can be related to the QES Coulomb and
the anharmonic oscillator potentials, respectively, provided suitable
limiting procedures are applied. It has been demonstrated that the Coulomb
correlation problem for a two-electron system in an external oscillator
potential corresponds to the particular case of the QES Coulomb potential.

We plan to extend the method to the solution of the multi-dimensional Schr%
\"{o}dinger equation. In particular, the Jahn-Teller problems can be solved
by using the method developed for quasi exactly solvable potentials.

\end{document}